\documentclass[amsmath,amsfonts,prl,aps,twocolumn,twoside,superscriptaddress]{revtex4}

\usepackage{epsfig,verbatim}
\usepackage{graphicx}
\usepackage{amsmath}
\usepackage{revsymb}

\def\squareforqed{\hbox{\rlap{$\sqcap$}$\sqcup$}}

\def\qed{\ifmmode\squareforqed\else{\unskip\nobreak\hfil

\penalty50\hskip1em\null\nobreak\hfil\squareforqed

\parfillskip=0pt\finalhyphendemerits=0\endgraf}\fi}

\def\endenv{\ifmmode\;\else{\unskip\nobreak\hfil

\penalty50\hskip1em\null\nobreak\hfil\;

\parfillskip=0pt\finalhyphendemerits=0\endgraf}\fi}

\mathchardef\ordinarycolon\mathcode`\:

\mathcode`\:=\string"8000

\def\vcentcolon{\mathrel{\mathop\ordinarycolon}}

\begingroup \catcode`\:=\active

  \lowercase{\endgroup

  \let :\vcentcolon

  }

\newcommand{\nc}{\newcommand}

\nc{\rnc}{\renewcommand}

\nc{\lbar}[1]{\overline{#1}}

\nc{\bra}[1]{\langle#1|}

\nc{\ket}[1]{|#1\rangle}

\nc{\ketbra}[2]{|#1\rangle\!\langle#2|}

\nc{\braket}[2]{\langle#1|#2\rangle}

\nc{\proj}[1]{| #1\rangle\!\langle #1 |}

\nc{\avg}[1]{\langle#1\rangle}

\nc{\Rank}{\operatorname{Rank}}

\nc{\smfrac}[2]{\mbox{$\frac{#1}{#2}$}}

\nc{\Tr}{\operatorname{Tr}}

\nc{\tr}{\operatorname{Tr}}

\nc{\id}{\operatorname{id}}

\nc{\1}{\openone}

\nc{\ox}{\otimes}

\nc{\dg}{\dagger}

\nc{\dn}{\downarrow}

\nc{\cA}{{\cal A}}

\nc{\cB}{{\cal B}}

\nc{\cC}{{\cal C}}

\nc{\cD}{{\cal D}}

\nc{\cE}{{\cal E}}

\nc{\cF}{{\cal F}}

\nc{\cG}{{\cal G}}

\nc{\cH}{{\cal H}}

\nc{\cI}{{\cal I}}

\nc{\cJ}{{\cal J}}

\nc{\cK}{{\cal K}}

\nc{\cL}{{\cal L}}

\nc{\cM}{{\cal M}}

\nc{\cN}{{\cal N}}

\nc{\cO}{{\cal O}}

\nc{\cP}{{\cal P}}

\nc{\cQ}{{\cal Q}}

\nc{\cR}{{\cal R}}

\nc{\cS}{{\cal S}}

\nc{\cT}{{\cal T}}

\nc{\cX}{{\cal X}}

\nc{\cY}{{\cal Y}}

\nc{\cZ}{{\cal Z}}

\nc{\supp}{{\operatorname{supp}}}

\nc{\var}{\operatorname{var}}

\nc{\rar}{\rightarrow}

\nc{\lrar}{\longrightarrow}

\nc{\polylog}{\operatorname{polylog}}

\def\e{\epsilon}

\nc{\RR}{{{\mathbb R}}}

\nc{\CC}{{{\mathbb C}}}

\nc{\FF}{{{\mathbb F}}}

\nc{\NN}{{{\mathbb N}}}

\nc{\ZZ}{{{\mathbb Z}}}

\nc{\PP}{{{\mathbb P}}}

\nc{\QQ}{{{\mathbb Q}}}

\nc{\UU}{{{\mathbb U}}}

\nc{\EE}{{{\mathbb E}}}

\nc{\Icoh}{{I^{\rm coh}}}

\nc{\Qca}{{Q_{\rm ss}}}

\nc{\Qcaa}{{Q^{(1)}_{\rm ss}}}

\nc{\Dcaa}{{D^{(1)}_{{\rm ss}\rightarrow}}}

\nc{\Dca}{{D_{{\rm ss}\rightarrow}}}

\nc{\be}{\begin{equation}}

\nc{\ee}{{\end{equation}}}

\nc{\bea}{\begin{eqnarray}}

\nc{\eea}{\end{eqnarray}}

\nc{\<}{\langle}

\rnc{\>}{\rangle}

\nc{\Hom}[2]{\mbox{Hom}(\CC^{#1},\CC^{#2})}

\nc{\rU}{\mbox{U}}

\begin{document}

\title{Can non-private channels transmit quantum information? }
\author{Graeme Smith}\email{gsbsmith@gmail.com}
\affiliation{IBM T.J. Watson Research Center, Yorktown Heights, NY
10598, USA}
\author{John A. Smolin}\email{smolin@watson.ibm.com}
\affiliation{IBM T.J. Watson Research Center, Yorktown Heights, NY
10598, USA}

\date{\today}

\begin{abstract}
We study the power of quantum channels with little or no capacity for
private communication.  Because privacy is a necessary condition for
quantum communication, one might expect that such channels would be of
little use for transmitting quantum states.  Nevertheless, we find
strong evidence that there are pairs of such channels that, when used
together, can transmit far more quantum information than the sum of
their individual private capacities.  Because quantum transmissions are 
necessarily private, this would imply a large violation of additivity 
for the private capacity.  Specifically, we present
channels which display either (1) A large joint quantum capacity but
very small individual private capacities or (2) a severe violation of
additivity for the Holevo information.
\end{abstract}
\maketitle

Shannon's information theory, which mathematically formalizes the
problem of communication in the presence of noise, underlies the
reliability of all modern communications technologies
\cite{Shannon48}.  The cornerstone of Shannon's theory is his capacity
formula, which gives an elegant expression quantifying the capability of
a communication channel for noiseless transmission.  Capacities
quantify the ultimate limits on communication with a physical channel,
and provide essential insight for the design of practical error
correction and mitigation schemes \cite{RU03}.

The starting point of information theory is to model the noise in a
communication link probabilistically.  In many physical systems this
is a reasonable approximation, as evidenced by the engineering success
of the theory.  However, the physical systems underlying all
communication are fundamentally quantum mechanical and when quantum
effects become prominent, classical probabilistic modeling will
provide a poor approximation.
One of the first quantitative investigations 
of this was the work of Holevo \cite{Holevo73}, who gave an upper
bound on the capacity of a noisy quantum channel for classical
communication.

In these early investigations, quantum effects were generally
considered to be a nuisance---quantum mechanics was a fundamental
source of noise that had to be dealt with to enable faithful
communication.  In contrast, in 1984 Bennett and Brassard suggested
\cite{BB84} that quantum effects might be {\em useful} for carrying
out communication and cryptographic tasks that are impossible in a
classical theory.  Specifically, they proposed a quantum method for
unconditionally secure key distribution and classical communication.
These ideas spawned a broad array of work on both the theory and
experiment of quantum key distribution, and there are many groups
worldwide working on practical implementations.  Indeed, quantum key
distribution is, and probably will remain for some time, the only
practical quantum information based technology.

Much as the classical capacity of a channel characterizes its
capability for noiseless classical communication, the {\em private
capacity} of a quantum channel tells us about a channel's capability
for communication that is secret from an eavesdropper.  More formally,
the classical capacity of a quantum channel $\cN$ is denoted by
$C(\cN)$, and is defined as the maximal number of bits per channel use
that can be sent with transmission errors vanishing in the asymptotic
limit. The private capacity has the additional constraint
that an eavesdropper with access to the environments of the channels used 
\footnote{Any channel $\cN$ can be expressed as an isometry followed
by a partial trace: $\cN(\rho) = \Tr_E U \rho U^\dg$, with
$U:A\rightarrow BE$ satisfying $U^\dg U = I$.  $E$ is referred to as
the environment of the channel, and the eavesdropper is given access
to the channel $\widehat{\cN}(\rho) = \Tr_B U\rho U^\dg$.} can
learn arbitrarily little about the key.

Unfortunately, unlike the classical capacity of a classical channel,
no simple characterization is known for either the classical or
private capacity of a quantum channel.  For example, the classical
capacity of a quantum channel is known \cite{SW97,Holevo98} to satisfy
\begin{equation}
C(\cN) = \lim_{n \rightarrow \infty}\frac{1}{n}\chi(\cN^{\ox n}),
\end{equation} 
where the {\em Holevo information} is defined as
\begin{equation}
\chi({\cal N}) = \max_{\cal E} \chi(\cN,\cE)
\end{equation}
with 
\begin{equation}
\chi(\cN,\cE) = S\left(\sum_i p_i {\cal N}(\rho_i)\right) - \sum_i p_i S({\cal N}(\rho_i)) 
\end{equation}
for ${\cal E}$ an ensemble $\{p_i,\rho_i\}$ of probabilities $p_i$ and
quantum states $\rho_i$ and $S(\rho)=- {\rm Tr} \rho \log \rho$ is the
von Neumann entropy.  For some special channels it is known that
$C(\cN) = \chi(\cN)$---in other words, the limit is unnecessary.
It has, however, recently been reported that there exist channels for which
this is not true \cite{H08}, though the violation is extremely small.
Similarly, the private capacity satisfies
\begin{equation}\label{Eq:RegularP}
\cP(\cN) = \lim_{n\rightarrow \infty} \frac{1}{n}\cP^{(1)}(\cN^{\ox n}),
\end{equation}
where the {\em private information} is defined as 
\begin{equation}
\cP^{(1)}(\cN) = \max_{\cE} \left(\chi(\cN,\cE) - \chi(\widehat{\cN},\cE)\right)
\end{equation}
where the complementary channel $\widehat{\cN}$ is defined below [18].
In this case, it is known that the limit in
Eq.~(\ref{Eq:RegularP}) cannot be removed in general \cite{SRS08},
even for some very natural qubit channels.

These difficulties in evaluating capacities are closely related to the
family of problems known as additivity problems.  A real function,
$f$, on the set of quantum channels is said to be additive if $f(\cN
\ox \cM) = f(\cN)+f(\cM)$.  Determining whether a given function is
additive is a problem that arises constantly in quantum information
science in a variety of very natural settings \cite{Shor04}.  For
example, if it were possible to show that $\chi$ is additive, we would
immediately be able to conclude that $C(\cN) = \chi(\cN)$.  Similarly,
the fact that the regularization in Eq.~(\ref{Eq:RegularP}) cannot be
removed is a consequence of the fact that $\cP^{(1)}$ is {\em not}
additive.  In the context of quantum Shannon theory, the importance of
additivity questions is twofold: first, showing additivity of some
entropic quantity may often lead to a simple capacity formula; and
second, when a capacity is additive it uniquely specifies the
channel's communication capabilities independent of what other
channels may be available.

The quantum capacity of a channel is the maximal rate, in qubits 
per channel use, at which a sender can reliably transmit quantum information
in the asymptotic limit.  The essential feature of the quantum capacity is that
transmission must be reliable not only on a set of orthogonal states, but also 
on arbitrary superpositions.  The quantum capacity of a channel $\cN$ is denoted $\cQ(\cN)$.  

It was recently shown that the capacity of a quantum channel for
quantum communication is not additive \cite{SY08}.  In fact, the
quantum capacity is very strongly nonadditive: there are pairs of
quantum channels $\cN$ and $\cA$, both with a quantum capacity of
zero, that nevertheless can be combined to achieve a positive
capacity: $\cQ(\cN) = \cQ(\cA) = 0$ but $\cQ(\cN \ox \cA)>0$, where
$\cQ$ is the quantum capacity.  This {\em superactivation} is not yet
completely understood, but from \cite{SY08} it seemed to be related to
the existence of channels, termed ``private Horodecki channels'', with
zero quantum capacity but positive private capacity
\cite{HHHO03,HPHH05}.  Indeed, at the heart of \cite{SY08} is an
argument showing there is an $\cA$ with $\cQ(\cA)=0$ such that if
$\cN$ has $\cQ(\cN)=0$ but $\cP(\cN)>0$ the joint capacity $\cQ(\cN\ox
\cA)\geq (1/2)\cP(\cN)$.  One interpretation of this effect is that,
while neither $\cN$ nor $\cA$ is capable of transmitting noiseless
quantum information, the two channels have complementary capabilities
for communication which can be combined for sending quantum
information.  Naturally, one would expect $\cN$'s capability is
somehow related to it's private capacity.

In this work we connect the additivity questions for the Holevo
information and the private capacity by showing that either one or the
other is highly nonadditive.  Specifically, we show that for every
$\e>0$ there is a family of channels $\cR_{d}^{\e}$ with increasing
input dimension $d$ and a channel $\cA$ with $\cP(\cA) = 0$ such that
either (1) $\cP(\cR^{\e}_d \ox \cA)$ is $O(\log d)$ larger than
$\cP(\cR^{\e}_d)$ or (2) $C(\cR^{\e}_d)$ is $O(\log d)$ larger than
$\chi(\cR^{\e}_d)$.  Assuming the additivity of $\chi$ for this channel, which we
regard as more likely, allows us to conclude that $\cP(\cR^\e_d) \leq
\e$ but $\cP(\cR^{\e}_d \ox \cA)\gtrsim \frac{1}{2}\log d$.

Thus, while it was natural to conjecture that ``privacy'' is the
feature that the private Horodecki channel contributes allowing the
superactivation effect, our results suggest the situation cannot be
quite as simple as that.  
Indeed, it appears that two channels with little or no private capacity
an be combined to send an arbitrarily large amount of private and even 
{\em quantum} data.

\begin{figure} 
\centering
\includegraphics[scale=.25]{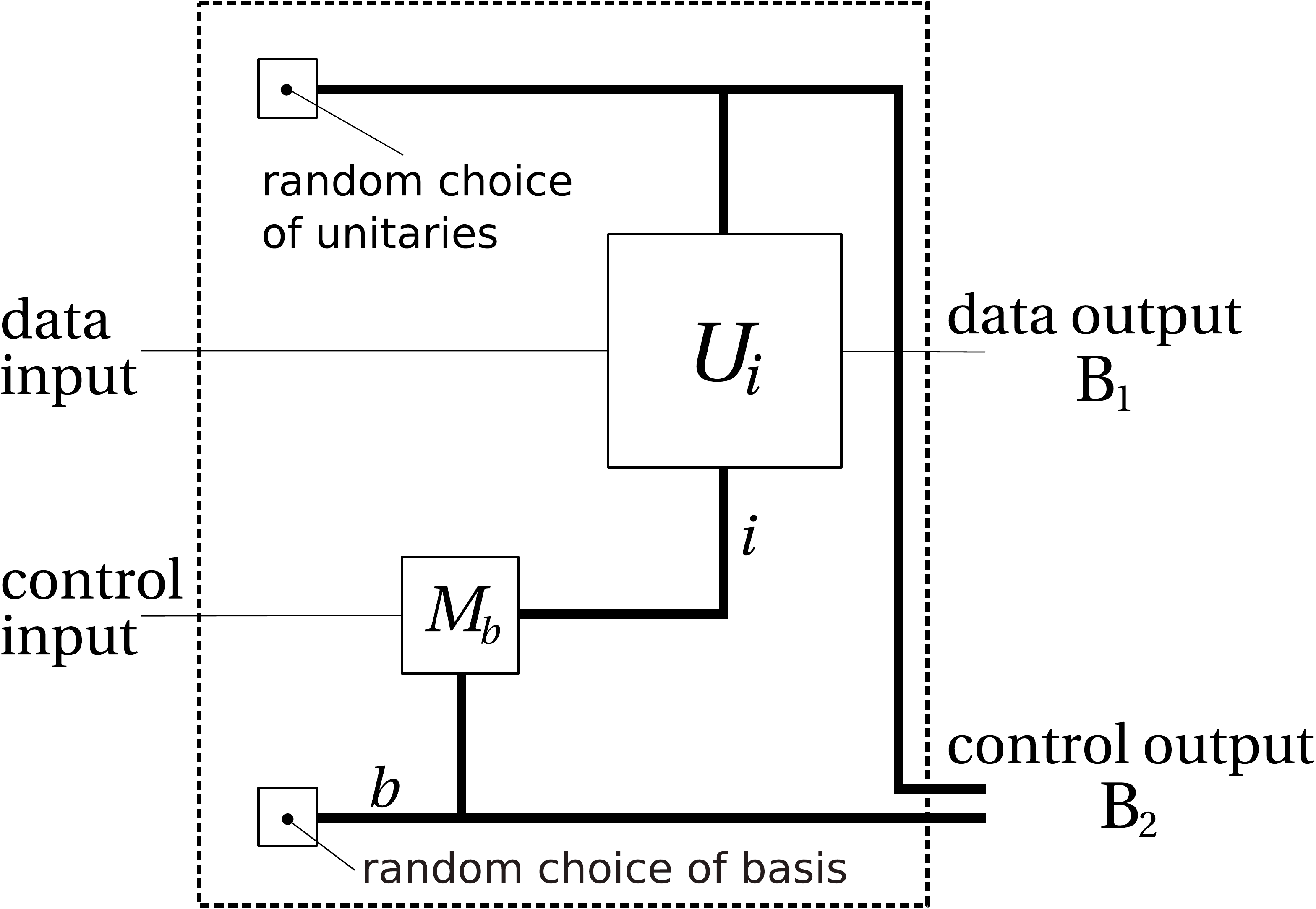}
\caption{Retro-correctable channel.  A retro-correctable channel
has two inputs and two outputs. Thin lines contain quantum data, while thick lines represent classical data. The control input is measured in a 
random basis.  The result of this measurement is used to select a member from
a random set of unitaries which is applied to the data input, which is
then outputted as the data output.  The control output contains the choice
of the set of random unitaries and the random basis.\label{Fig1}}
\end{figure}

Our main building block in what follows will be the retro- or echo-correctable
channels of \cite{BDSS06} (see FIG. \ref{Fig1}).  The standard echo-correctable
channel ${\cal R}_{d}^{\e}$ has a $d$-dimensional data input and a
corresponding output; a control input of dimension $c = (K/\e^2)d(\log
d)^4$ \footnote{The alert reader will note the additional power of
$\log d$ beyond that needed in \cite{BDSS06}.  This is the result
of a slightly more conservative approach we have taken and, being a
sublinear factor, doesn't change any capacity quantity or qualitative
result.}  with $K$ a constant; and an infinite-dimensional classical
control output. The channel internally selects a random basis $b$, for
${\cal H}_{c}$, and a set of $c$ random unitaries $\{U\}=U_1...U_c$ on
${\cal H}_d$. The channel measures the control input in the basis $b$,
yielding result $j\in\{1...c\}$ and according to that result applies
one of the unitaries $U_i$ to the data input, which is then emitted as
the data output $B_1$. The channel also emits a classical control output $B_2$
consisting of the random basis $b$ and the set of random unitaries
$\{U\}=U_1...U_c$.  It does not, however, emit the measurement result
$i$ but keeps it hidden.

It can be shown \cite{BDSS06} that for any $\e>0$ and
sufficiently large $d$ that $\chi(\cR^\e_d) \leq \e$.  Thus, if
$C=\chi$ such a channel has almost no classical capacity and since the
classical capacity upper bounds $\cP$, it too becomes small.

However, when used in combination with an erasure channel $\cA_e^p$ which takes
a $c$-dimensional input and with probability $1-p$ transmits the input to the
output perfectly, but with probability $p$ outputs only an erasure flag, then
the combination has a great deal of both quantum and private capacity:  
$\cP \ge \cQ \simeq (1-p) \log d$.  This is most striking, of course, when 
$p\ge 1/2$ since it is then that the erasure channel
has no private or quantum capacity at all \cite{BDS97}.

\begin{figure}
\centering
\includegraphics[scale=.45]{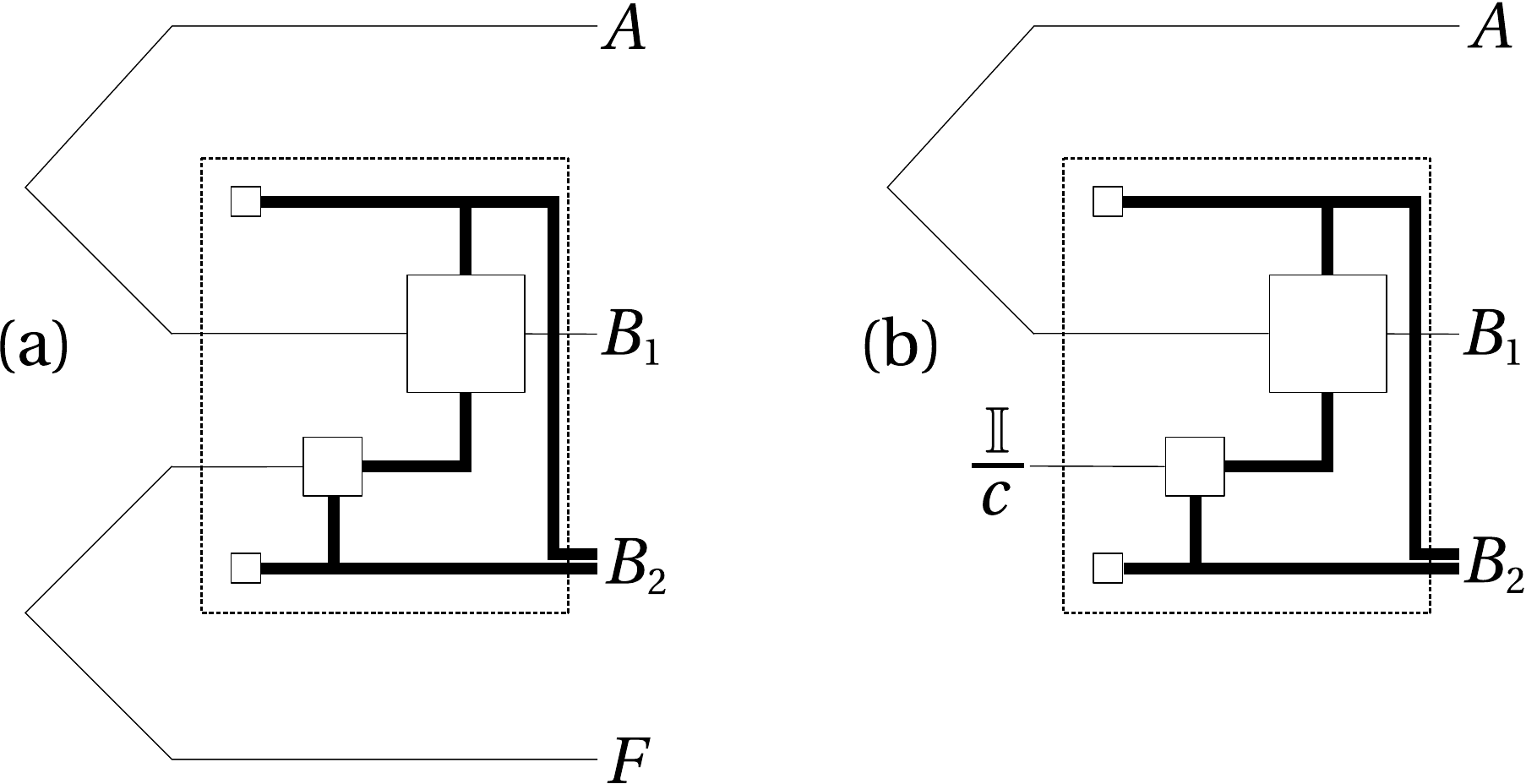}
\caption{Evaluating joint coherent information when Alice feeds half of 
one maximally entangled state into the data input of a retro-correctable 
channel, keeping its purification $A$, and feeds another maximally entangled 
state into the control input and a 50\% erasure channel.  (a) When the
erasure channel doesn't erase, Bob receives the purification $F$ of the control
input.  (b)  When the erasure channel erases, it is as if a maximally mixed
state $\frac{\mathbb{I}}{c}$ was fed into the control input.\label{Fig2}}

\end{figure}

The way to use the two channels together is shown in FIG. \ref{Fig2}. Alice prepares
a maximally-entangled state of $d \times d$-dimensions on Hilbert
space $AA'$ and another of $c \times c$-dimensions on space $FF'$.
She feeds the $A'$ and $F'$ systems into the data and control inputs
of $\cR^\e_d$ respectively, and she also puts the $F$ system into
the erasure channel, whose output we will call $B_3$.  

The coherent information of the resulting bipartite state
$\rho_{A:B_1B_2B_3}$ is a lower bound on $\cQ$ and
$\cP$\cite{D03,Shor02,Lloyd97}:
\begin{equation}\label{Eq:CohInfCond}
I_{\rm coh}  = S(B_1 B_3|B_2) - S(AB_1 B_3|B_2)\ .
\end{equation}
Here, since $B_2$ is classical, the conditional entropies
are given by averages over $b_2$, the possible values of $B_2$:
\begin{eqnarray}
\nonumber &S(B_1B_3|B_2) = \int {\rm d}b_2 S(\rho^{b_2}_{B_1B_3})\\
\nonumber &S(AB_1B_3|B_2) = \int {\rm d}b_2 S(\rho^{b_2}_{AB_1B_3})
\end{eqnarray}
 where
$\rho^{b_2}_{AB_1B_3}$ and $\rho^{b_2}_{B_1B_3}$ are conditional
states given $b_2$.  Using the slightly nonstandard expression in
Eq.~(\ref{Eq:CohInfCond}) allows us to avoid any complications due to
the fact that $B_2$ is infinite dimensional.

The coherent information is straightforward to calculate
since the erasure channel's flag breaks the quantity into the sum of two terms:
\begin{equation}
I_{\rm coh}=(1-p) I_{\rm coh}^{\rm not\ erased}  + p I_{\rm coh}^{\rm erased} .
\end{equation}

In the unerased case, since Bob knows what basis to measure in he can
measure the $F$ system which he has received through the successful
use of the erasure channel and determine exactly which $U_i$ has
occurred.  Thus
\begin{equation}
I_{\rm coh}^{\rm not \ erased}= \log d\ .
\end{equation}
When the $F$ system {\em is} erased, the the conditional entropy of the $AB_1$ given $B_2$
is at most $\log c= \log d + 4 \log \log d + \log(K/\e^2)$,
while the conditional entropy of $B_1$ given $B_2$ is $\log d$.  So, we have
\begin{equation}
I_{\rm coh}^{\rm  erased}\ge -4\log \log d - \log(K/\e^2)
\end{equation}
and
\begin{equation}
I_{\rm coh} \ge (1-p) \log d - p(4\log\log d+ \log (K/\e^2) ),
\end{equation}
which is positive as $d\rightarrow \infty$ for 
\begin{equation}
\frac{1-p}{p} > \frac{4 \log \log d + \log(K/\e^2)}{\log d}.
\end{equation}

\begin{figure}
\centering \includegraphics[scale=.42]{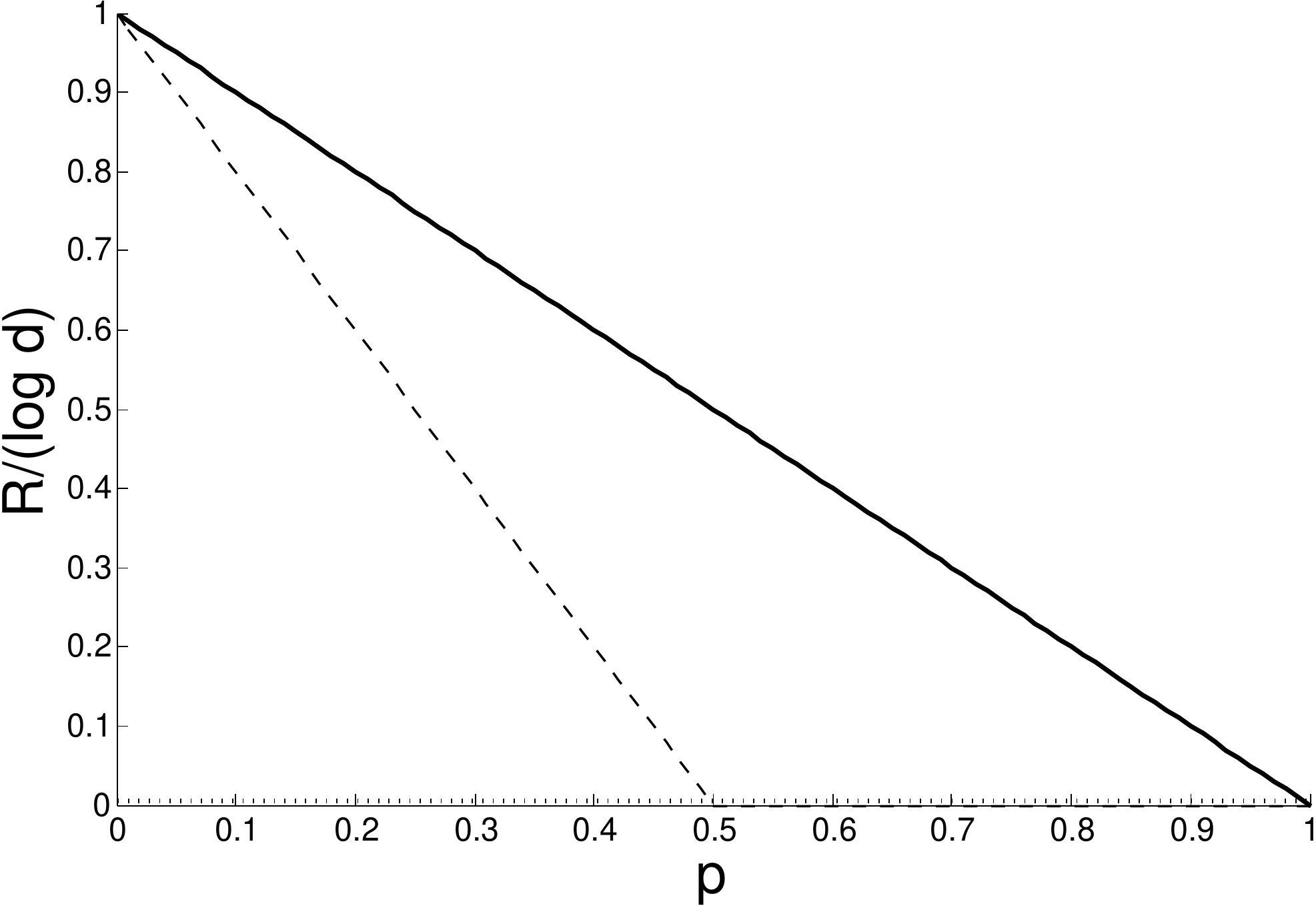}
\caption{Joint and individual private capacities of $\cR^\e_d$ and
$\cA_e^p$, normalized by $\log d$ in the $d\rightarrow \infty$ limit.
The solid line is the achievable rate using $\cR^\e_d$ with $\cA_e^p$
and the protocol described in the text.  The dashed line is the
capacity of $\cA_e^p$ alone.  The dotted line is an upper bound on the
capacity of $\cR^\e_d$ assuming additivity of $\chi$.\label{Fig3}}
\end{figure}

A consequence of the above argument is that at least one of $\cP$ and
$\chi$ violates additivity severely.  In particular, letting $p=1/2$,
if $\cP(\cR^{\e}_d \ox \cA_e)- \cP(\cR^\e_d) = o(\log d)$ we have
$\cP(\cR^\e_d) = O(\log d)$.  Since $C \geq \cP$, this implies that
$C(\cR^\e_d) = O(\log d)$ while $\chi(\cR^\e_d) \leq \e$.  Otherwise,
if $\cP(\cR^{\e}_d \ox \cA_e)- \cP(\cR^\e_d) = O(\log d)$, we have a
large violation of additivity for $\cP$, since then $\cP(\cA_e)=0$,
but $\cP(\cR^\e_d \ox \cA_e) \gg \cP(\cR^\e_d)$. We have plotted the joint
and individual capacities or $\cR_d^\epsilon$ (assuming it has
additive $\chi$) and $\cA_e^p$ in FIG. \ref{Fig3}.

In summary, we have explored violations of additivity arising from two channels, 
$\cR^{\e}_d$ and $\cA^{1/2}_e$.  $\cR^\e_d$ is a retro-correctable channel, described in 
FIG 1, and satisfies $\chi(\cR^\e_d) \leq \e$.  $\cA^{1/2}_e$ is a $50\%$ quantum erasure 
channel, whose private capacity is zero.  Our main result, illustrated in FIG 2, is that 
these two channels can be used together to transmit large amounts of quantum information.
This leaves only two possibilities: either (1) $\cR^\e_d$ has a large classical capacity, which
would imply severe nonadditivity of $\chi$ or (2) a large violation of additivity for the 
private capacity.   
    
As we have mentioned above, we consider the extreme nonadditivity of
$\chi$ for the retro-correctable channel to be rather unlikely, and
tend to believe instead that it is $\cP$ which is nonadditive.  We
believe this despite the recent results of Hastings \cite{H08} since
his results are a tiny effect for a very specially designed family of
channels.  It is nevertheless an important open problem to find an
argument that shows nonadditivity of the private capacity without
additivity assumptions on $\chi$.  While the $p\geq 1/2$ erasure
channels we have used have quantum and private capacities exactly
equal to zero, we have only been able to show that the
retro-correctable channels, $\cR^\e_d$, have capacity less than $\e$
(even this is conditional on the additivity of $\chi$).  One would
hope for the stronger result of pairs of channels with strictly zero
private capacity that can jointly allow nonzero private capacity.
This would be parallel to the quantum capacity findings in
\cite{SY08}. In the quantum setting, there are two distinct types of
zero capacity channels---PPT channels and channels whose
environment can simulate the channel output (sometimes called
``antidegradable'').  Unfortunately, the only type of channels known to
have zero private capacity are the antidegradable ones (which include
symmetric channels as a special case).  Because the product of
antidegradable channels is itself antidegradable, and therefore has
zero private capacity, a necessary step for finding genuine
superactivation would be identifying a class of channels with zero
private capacity that are not antidegradable.  Finding such channels
is an intriguing open problem.

We do know that the classical capacity is often a very weak bound on
the private capacity (for example, a classical channel always has
exactly zero private capacity, regardless of its classical capacity).
It is then plausible that the private capacity of $\cR_d^\epsilon$, or if not
that $\cR_d^\epsilon$ with some small additional noise that would leave the
joint capacity essentially unchanged,
may actually be zero.   If this is so, then the superactivation effect
of \cite{SY08} requires no privacy and we are left to wonder just what
it is that $\cR_d^\epsilon$ provides.

\bibliographystyle{apsrev}


\end{document}